\documentclass[aps,prc,preprint,groupedaddress,figure]{revtex4-1}
\usepackage{graphicx}

\newcommand{\nonubb}  {0 \nu \beta \beta}

\newcommand{\MJ}      {{\sc{Majo\-ra\-na}}}
\newcommand{\DEM}	{{\sc{Dem\-on\-strat\-or}}}
\newcommand{\GERDA} {GERDA}

\newcommand{\cogent} {CoGeNT}
\newcommand{\munu} {$\mu_\nu$}

\def\nuc#1#2{${}^{#1}$#2}

\def\ppc{P-PC}                          

\begin{document}

\title{Neutrino physics with an intense \nuc{51}{Cr }source and an array of
low-energy threshold HPGe detectors.}
\author{Reyco Henning}
\email[]{rhenning@physics.unc.edu}
\affiliation{University of North Carolina at Chapel Hill and Triangle Universities Nuclear Laboratory}

\date{\today}

\begin{abstract}
We study some of the physics potential of an intense  $1\,\mathrm{MCi}$ \nuc{51}{Cr} source combined with the  \MJ\ \DEM\ enriched germanium detector array. The \DEM\ will consist of detectors with ultra-low radioactive backgrounds and extremely low energy thresholds of~$\sim 400\,\mathrm{eV}$. We show that it can improve the current limit on the neutrino magnetic dipole moment. We briefly discuss physics applications of the  charged-current reaction of the \nuc{51}{Cr} neutrino with the \nuc{73}{Ge} isotope. Finally,  we argue that the rate from a realistic, intense tritium source is below the detectable limit of even a tonne-scale germanium experiment.
\end{abstract}


\maketitle

\section{Introduction}
\label{se:introduction}

High-purity Germanium (HPGe) Detectors have been used extensively in the detection of ionizing radiation, especially in applications that require good energy resolution. Recent work has demonstrated that HPGe crystals with a p-type point-contact (\ppc ) geometry can have sub-keV energy thresholds and applications in particle astrophysics~\cite{barb07}. The \cogent\ collaboration has demonstrated the competitive dark matter sensitivity of this technology and an extremely low-energy threshold of $400\,\mathrm{eV}$~\cite{aals08, aals10}. In ref.~\cite{aals10} they also provide a spectrum of a P-PC crystal detector operated in an ultra-low background cryostat at a depth of \mbox{$2100\,\mathrm{m.w.e}$}. Currently, the \MJ\ ~\cite{henn09} and \GERDA\ ~\cite{abt04} collaborations are constructing experiments that will deploy arrays of \ppc\  HPGe detectors in ultra-low background configurations. 
\\
This paper discusses the potential of the \MJ\ \DEM\ experiment to improve limits on the magnetic dipole moment of the neutrino (\munu ) by searching for neutrino scattering from an intense \nuc{51}{Cr} source off electrons in germanium. We use the current measurement of the background spectrum from the \cogent\ experiment, extrapolate it to the case of the \MJ\ \DEM , and estimate the sensitivity of the \MJ\ \DEM . We also discuss applications of the  charged-current reaction of the \nuc{51}{Cr} neutrino with the \nuc{73}{Ge} isotope and how it relates to neutrino oscillation measurements.
\\
The \MJ\ \DEM\ ~\cite{henn09} is a modular instrument composed of two to three cryostats built from ultrapure electroformed copper, each containing 20 kg of 0.6 kg \ppc\ HPGe detectors.
Its main goal is to search for the neutrinoless double-beta decay ($\nonubb $-decay)  of the \nuc{76}{Ge} isotope. It will also be sensitive to light WIMP dark matter in the $1-10\,\mathrm{GeV/c^2}$ mass range. About one half of the detectors will be manufactured from isotopically enriched germanium, resulting in a \nuc{76}{Ge} mass of up to 30 kg. The array requires extensive shielding from external radiation sources. Its shield will consist of different layers that consists of (from inside to outside) electroformed and commercial high-purity copper, high purity lead, a radon exclusion box, an active muon veto and finally a layer of neutron moderator. The experiment will be located in a clean room at the 4850~foot level of the Sanford Underground Laboratory in Lead, South Dakota. The first module of the \DEM\ will be deployed underground at the Sanford Laboratory in 2012.  
\\
The concept of high activity radionuclide neutrino sources is not new. A \nuc{51}{Cr} source was first suggested in 1978 by Raghavan~\cite{ragh78}, and \nuc{51}{Cr} sources of megacurie (MCi) intensities were used by the SAGE~\cite{abdu96, abdu99} and GALLEX~\cite{crib96} experiments in the 1990's. Other authors have consider the use of such a \nuc{51}{Cr} source with indium-loaded organic liquid scintillator~\cite{grie07} and liquid noble gas based detectors~\cite{link10} to study neutrino scattering. \ppc\ detectors are also good candidates for measuring neutrino scattering from this source. Their relatively small size allows them to be located close to the source, significantly enhancing the neutrino flux. In fact, the original motivation of the \cogent\ experiment was the search for coherent neutrino-nuclear scattering using reactor neutrinos~\cite{barb07}. 

\section{Theoretical Motivation and Background}
\label{theoretical_motivation}

We will consider neutrino scattering from atomic electrons. The electromagnetic interaction between the putative neutrino magnetic dipole moment and the electron contributes to this process. The SM with an extension to include massive Dirac neutrinos gives the neutrino a magnetic dipole moment via radiative corrections as~\cite{voge89} :

\begin{equation}
\label{eq:SMmunu}
\mu_\nu =  \frac{3 G_F m_e m_\nu}{4 \sqrt{2}\pi^2} = 3.2 \times 10^{-19}\left(\frac{m_\nu}{1\,\mathrm{eV}}\right)\mu_B
\end{equation}

where $m_e$ and $m_\nu$ are the electron and neutrino masses respectively, $G_F$ is the Fermi constant, and $\mu_B$ is the Bohr magneton. This value is quite small and even the most general estimates for Dirac neutrinos place an upper limit of $\mu_\nu < 10^{-14}\mu_B$~\cite{bell05}. However, models exist with Majorana neutrinos that potentially have larger magnetic dipole moments. These models are already constrained by the current best experimental limits. Hence, a measurement of $\mu_\nu > 10^{-14}\mu_B$ would imply that the neutrino is Majorana and also hint at physics at the TeV scale or beyond~\cite{bell06}.  
\\
The neutrino free-electron scattering cross-section due to the magnetic dipole moment is given as~\cite{voge89}:

\begin{equation}
\label{eq:free_e_dipole_xsect}
\frac{d\sigma}{dT} = \frac{\pi \alpha^2 \mu_\nu^2}{m_e^2}\frac{1-T/E_\nu}{T}
\end{equation}

where $T$ is the electron recoil energy, $\alpha$ the fine structure constant, $m_e$ the mass of the electron, and $E_\nu$ the energy of the incident neutrino. 
A recent study that included the effects of coherent scattering of electron neutrinos from atomic electrons modifies the cross-secttion at low energies to~\cite{wong10} :

\begin{equation}
\label{eq:atomic_e_dipole_xsect}
\frac{d\sigma}{dT} \simeq \mu_\nu^2 \frac{\alpha}{\pi}(\frac{E_\nu}{m_e})^2\frac{1}{T}\sigma_{\gamma A}(E_\gamma = T)
\end{equation}

where $\sigma_{\gamma A}(E)$ is the photo-electric scattering cross-section for gamma-rays of energy $E$ from atom $A$. 
This cross-section has a significant enhancement at low recoil energies over the free-electron scattering cross-section. This further motivates the use of detectors with very low energy thresholds. However, the cross-section from these authors is in contradiction with other studies that found a slight reduction in the cross-section~\cite{faya92} or that claim that coherence effects are not applicable~\cite{goun02, volo10}. Given this theoretical uncertainty, we will consider both free electron and coherent atomic scattering in this paper. 
\\
The GEMMA collaboration has recently posted the best current limits for $\mu_\nu$~\cite{bena10}. With the atomic enhancement to the cross-section from ref.~\cite{wong10} they claim a limit of $\mu_\nu < 5.0\times 10^{-12}\mu_B$. They obtained a limit of $\mu_\nu < 3.2\times 10^{-11}\mu_B$ assuming free-electron scattering. We will use both these limits in our analysis and figures. More stringent, but model-dependent, limits have been derived from astrophysical processes. The energy loss due to plasmon decay into neutrinos in globular cluster red giant stars were used to place a limit of $\mu_\nu < 3\times10^{-12}\mu_B$ for both Majorana and Dirac neutrinos~\cite{raff99}. A more stringent limit for Dirac neutrinos of $(1.1-2.7)\times10^{-12}\mu_B$ is obtained using data from SN1987A~\cite{kuzn09}.

\section{\nuc{51}{Cr} Source}
\label{se:source}

\nuc{51}{Cr} decays via electron-capture (EC) into \nuc{51}{V} and emits a neutrino with several possible discrete energies and branching ratios as shown in Table~\ref{tab:cr51_modes}. The dominant mode (90\%) is into the ground state of \nuc{51}{V} with the emission of a neutrino around $747\,\mathrm{keV}$. The exact energy depends on the atomic shell from which the electron was captured. A second mode (10\%) into an excited state of \nuc{51}{V} emits a neutrino around $426\,\mathrm{keV}$. The nucleus subsequently immediately de-excites with the emission of a $320\,\mathrm{keV}$ gamma. 
There is also a small branching ratio ($10^{-4}$) for inner-bremsstrahlung emission.
\\
For this paper we consider an intense \nuc{51}{Cr} source primarily because it has been demonstrated that one can make sources from this isotope at $1\,\mathrm{MCi}$ intensities. It also emits a nearly mono-energetic neutrino making data analysis easier. The $320\,\mathrm{keV}$ gamma can easily be shielded, and the relatively short half-life of 27.7~days makes the source safe after a few months. Other authors have also considered the EC-decay isotopes \nuc{65}{Zn}~\cite{alva73}, \nuc{37}{Ar}~\cite{haxt88}, and \nuc{152}{Eu}~\cite{crib96}. The authors from~\cite{crib96} claim that \nuc{51}{Cr} was the best option for practical and economic reasons, as well as its lack of higher energy gamma-rays that would make handling difficult. 

\begin{table}
\begin{tabular}{|c|c|}
\hline \hline
\textbf{Neutrino Energy (keV)} & \textbf{Branching Ratio} \\
\hline
751.9 & 9.0\% \\
746.6 & 81.0\% \\
431.8 & 1.0\% \\
426.5 & 9.0\% \\
\hline\hline
\end{tabular}
\caption{\label{tab:cr51_modes} Energies and branching ratios of neutrinos emitted during \nuc{51}{Cr} decay. Electron capture from M-shells and higher were ignored. Based on data taken from~\cite{nudat06, abdu96}.}
\end{table}

\nuc{51}{Cr} is manufactured by exposing chromium that has been enriched in \nuc{50}{Cr} to a high neutron flux in a nuclear reactor core. Activities directly after exposure of $1.7\,\mathrm{MCi}$~\cite{crib96} and $517\,\mathrm{kCi}$~\cite{abdu99} have been achieved. However, the manufacturing of these sources is technically challenging. Only a few reactors in the world have the flexibility and short cycle to allow the reconfiguration of the core that is required, and power reactors are generally not suitable. Great care must be taken in the production of the enriched chromium to avoid the introduction of contaminants, since these may become activated under neutron bombardment and cause the source to emit high energy gamma-rays that cannot be effectively shielded. The source itself has to be shielded from the internal $320\,\mathrm{keV}$ gamma-rays, typically with tungsten. It must also be cooled to remove the thermal heat from the radionuclide decay. Finally, careful measurements, preferably using more than one method, must be done to ascertain the final activity of the source. The reader is referred to references~\cite{crib96, abdu99} for a detailed discussion of the technical aspects of the source production. Despite these challenges, there does exist interest and capability for the production of such sources in Russia at an approximate cost of ten million US dollars per source~\cite{gavr10}. Radiochemical experiments with gallium have also recently been proposed to search for sterile neutrino oscillations using such a source~\cite{gavr10a}. There are also formidable political and logistical challenges to rapidly transporting such a source from Russia to the United States, likely requiring the involvement of high-level administrators in both governments. 
\\
Another interesting source isotope to consider for P-PC detectors is tritium.  Tritium undergoes beta-decay with an end-point of $18.6\,\mathrm{keV}$ and emits a continuum of neutrinos out to its end-point. Such low energy neutrinos have interesting applications and experiments utilizing tritium neutrino sources have been proposed~\cite{nega01, kope03, mcla04, giom04}. However, a tonne-scale \MJ\ combined with the most intense tritium source ($40\,\mathrm{MCi}$) that has been realistically proposed~\cite{bogd10} would only observe a few scatters per year at very low energies. This is at least a few orders of magnitude below the most optimistic background projections (see below). At this time a tritium source program for \MJ\ does not appear viable. 

\section{Background Assumptions}
\label{se:background_assumptions}

The ultimate sensitivity to $\mu_\nu$ of the \MJ\ experiment is determined by its backgrounds at low energies. Primordial uranium and thorium decay chain daughters, cosmic-rays, and cosmic-ray induced radioactive isotopes contribute to the background. In the energy range of interest here, ($E_r < 2\,\mathrm{keV}$), the spectrum has three main components: a flat continuum, an exponential rise at low energy, and two peaks due L-shell EC decays of \nuc{65}{Zn} and \nuc{68}{Ge} inside the germanium, as was shown by the \cogent\ experiment~\cite{aals10, mari10}.  
We use this information to estimate the background in an array of P-PC HPGe detectors, specifically the \MJ\ \DEM , in the range $0.5-2.0\,\mathrm{keV}$. We assume that we will perform a one month run with a \nuc{51}{Cr} source that has an average activity of $1\,\mathrm{MCi}$. This run will be performed after the initial 4~year physics run that will search for $\nonubb $-decay in the \DEM . There are two main benefits to this approach. It allows the cosmogenic \nuc{65}{Zn} and \nuc{68}{Ge} activity to die away, and it provides ample time for the collaboration to accurately measure and characterize the background at these low energies, allowing an accurate comparison between the spectra with and without the source.  
\\
The origin of the flat continuum in \cogent\ is unknown, but we assume that it consists of two components. The first is events near the deadlayers of the crystal that suffer incomplete charge collection, and the second is a low energy tail from all the high-energy sources. The \DEM\ will use cleaner detector construction materials and procedures that will significantly reduce the radioactive backgrounds present in the \cogent\ cryostat, possibly by as much as a factor~100. Events that suffer incomplete charge collection can be further mitigated using pulse-shape analysis. 
\\
The origin of the exponential rise at low energy in the \cogent\ data is also  unknown. One possibility mentioned by the authors of~\cite{aals10} is light WIMP dark matter. If this were the case, then obviously this rise will remain in the \DEM . However, the analysis of pulse-shapes at such low-energies is challenging, and it is difficult the quantify the efficacy of cuts to remove backgrounds. Another possibility is that the events in the rise are due to unknown tails in cuts on distributions that rely on the pulse shape discrimination applied in ref.~\cite{aals10}. This is currently an active area of R\&D within the \MJ\ collaboration.  These events may also be related to higher energy events with incomplete charge collection. If this is the case, then the reduction in radioactive backgrounds in the \DEM\ will lead to corresponding reduction in this rise as well. 
\\
These assumptions lead the authors to make the following quantitative, subjectively conservative, estimates:
\begin{enumerate}
\item The cosmogenic activity of \nuc{68}{Ge} and \nuc{65}{Zn} is reduced by a factor~50 during the four years underground since these isotopes have half-lives of 271 and 244~days respectively~\cite{nudat06}.  
\item We estimate that the background in the continuum is the \DEM\ is a factor~10 less than that observed in \cogent .
\item  We estimate that the low energy rise can be suppressed by a factor~10 in the \DEM\ relative to \cogent .
\end{enumerate}

Nuclear recoils from coherent neutrino-nuclear scattering for the \nuc{51}{Cr} neutrinos have energies that are too low ($15\,\mathrm{eV}$ before quenching) to be detected and do not form part of the background. Tree-level weak neutrino-electron scattering also does not contribute significantly at these source intensities and energies.

\section{Sensitivity of the \MJ\ \DEM\ }
\label{se:sensitivity}

\begin{figure}
\includegraphics[width=6in]{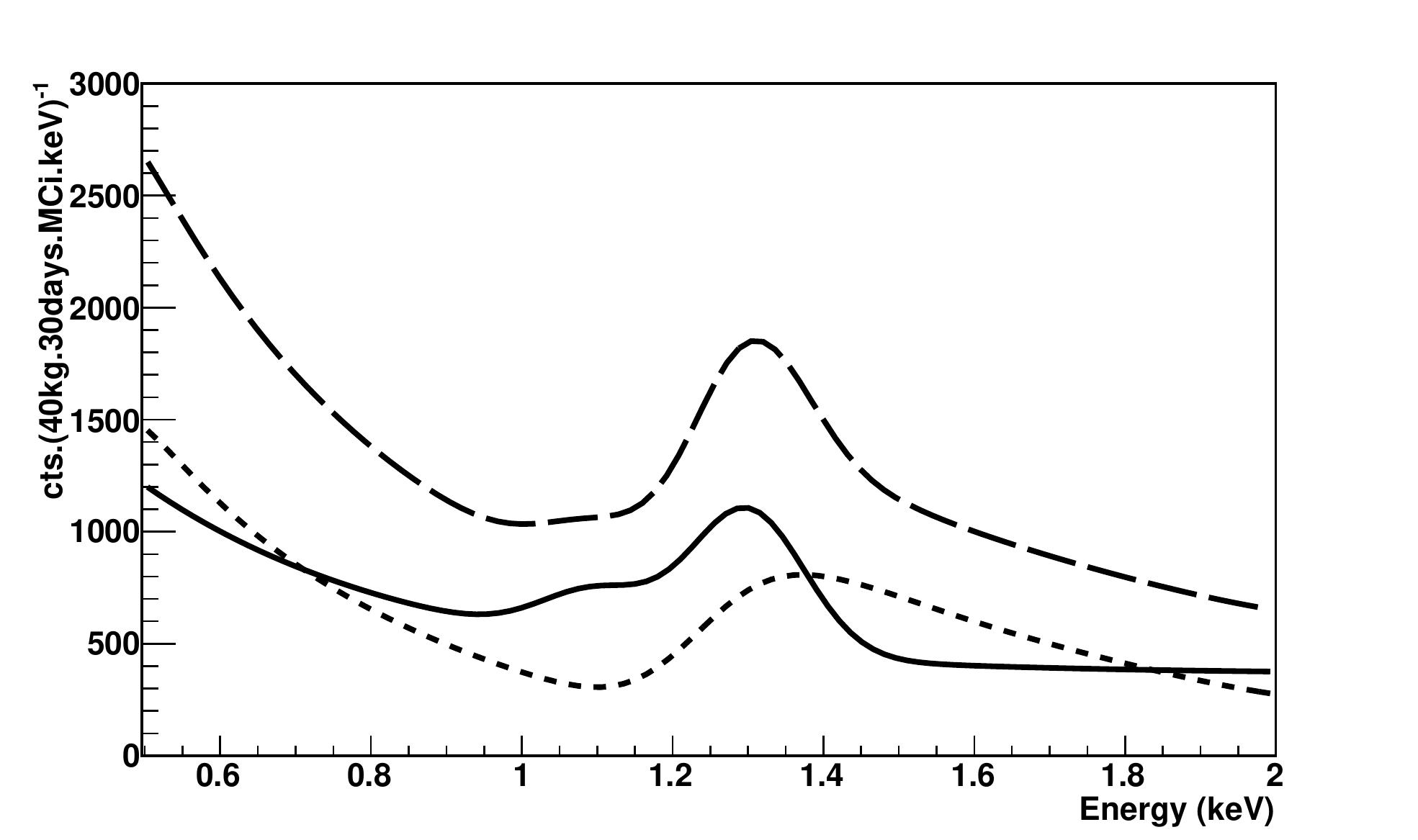}
\caption{\label{fig:spectraWWong} Shown are three spectra. The solid line is the \DEM\ background under the assumptions from the text. The dotted line is the expected signal from the \nuc{51}{Cr} source using the atomic correction from~\cite{wong10} and a value of $\mu_\nu = 5\times10^{-12}\mu_B$. The dashed line is the sum of the first two, ie. the spectrum the detector will measure in the presence of the source. The units on the y-axis correspond to the mass of the target Ge ($40\,\mathrm{kg}$), exposure time (30 days), and time-averaged source activity ($1\,\mathrm{MCi}$).}
\end{figure}

\begin{figure}
\includegraphics[width=6in]{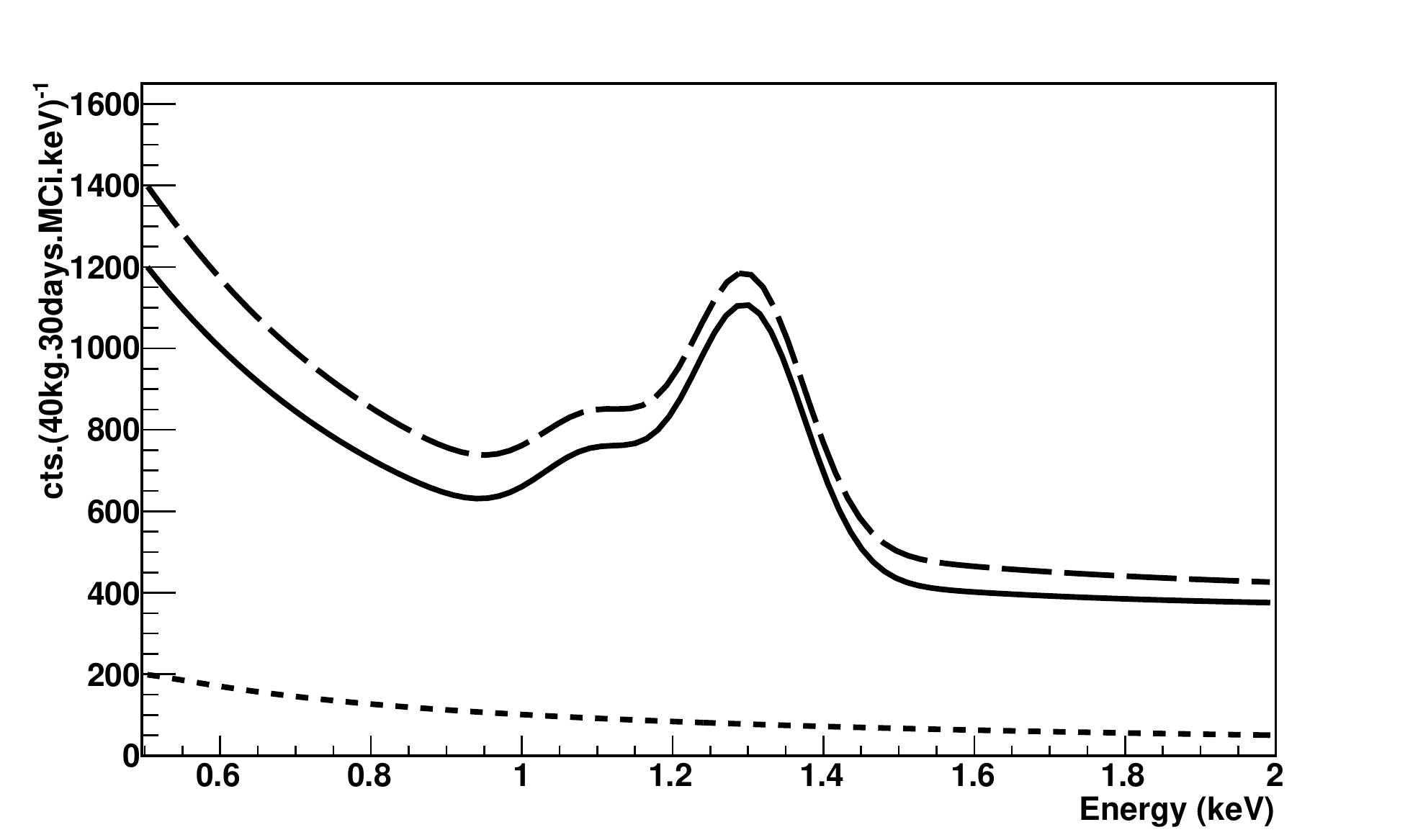}
\caption{\label{fig:spectraWOWong} Same as figure~\ref{fig:spectraWWong}, but the spectra from the \nuc{51}{Cr} source now uses the free electron scattering cross section and $\mu_\nu=3.2\times10^{-11}\mu_B$.}
\end{figure}

Given the background assumptions from section~\ref{se:background_assumptions}, we can generate an anticipated spectrum for two modules of the \MJ\ \DEM\ as shown in figures~\ref{fig:spectraWWong} and~\ref{fig:spectraWOWong}, with and without the presence of a \nuc{51}{Cr} source. In these figures and the subsequent analyses, we assume that the source is located at an average distance of $50\,\mathrm{cm}$ from the detector array. This places the source just outside the lead shield that will then shield the array from any spurious activity from the source. The spectra also include the finite energy resolution from the detectors.

The scattering cross sections for both the atomic and free electron cases are proportional to $\mu_\nu^2$ (eqns.~\ref{eq:free_e_dipole_xsect}~and~\ref{eq:atomic_e_dipole_xsect}), hence the expected recoil event rate scales directly with $\mu_\nu^2$. We have determined the sensitivity ($3 \sigma$ over background) of the \MJ\ \DEM\ to be as summarized in table~\ref{tab:money_table}. We have considered both the cases with and without atomic enhancement from~\cite{wong10}. We have also included an estimate for a 1-tonne germanium experiment for completeness, also assuming that the source is located at a average distance of $50\,\mathrm{cm}$ from the crystals in the array. In the 1-tonne case we simply scaled the detector mass and background accordingly and made no additional assumptions on further background reduction. This includes assuming the same 4~year period for the cosmogenic activity to die away.  It is clear that both the \DEM\ and tonne-scale experiment will be able improve the existing limits. The tonne-scale does not improve the sensitivity much, since it scales as the $\frac{1}{4}$ power of the detector mass (or exposure time) in the background-limited case. 

\begin{table}
\begin{tabular}{||c|c|c||}
\hline \hline
\textbf{Experiment} & \textbf{With Atomic Effects} & \textbf{Free-electron Scattering} \\
\hline \hline
\DEM\ & $1.6\times 10^{-12}$ & $2.7 \times 10^{-11}$ \\
\hline
1-tonne & $7\times 10^{-13}$ & $1.2 \times 10^{-11}$ \\
\hline \hline
\end{tabular}
\caption{\label{tab:money_table} Estimated sensitivities to \munu\ comparing different experimental configurations of the \MJ\ experiment with a $1\,\mathrm{MCi}$ \nuc{51}{Cr} source. The different calculations for the interaction cross-sections are also compared. Units are in Bohr magnetons. `With Atomic Effects' refers to the cross-section from~\cite{wong10}.}
\end{table}

The authors feel that the analysis and assumptions presented here are conservative. Further improvements in the sensitivity of the array can be achieved by using a more intense source, using multiple month-long runs with multiple sources, and performing a spectral shape analysis and not just a simple counting measurement. For a tonne-scale experiment, one can also consider placing the source in the middle of the array, significantly increasing the neutrino flux and senstivity. This analysis also assumed all natural germanium (\nuc{nat}{Ge}) detectors. Later modules in the \DEM\ will consist of enriched germanium (\nuc{enr}{Ge}) crystals. \nuc{enr}{Ge} has a significantly lower cosmic-ray activation rate than \nuc{nat}{Ge}, leading to further reduction in the background due to \nuc{65}{Zn} and \nuc{68}{Ge} decays.

\section{Charged Current Interaction}
\label{se:CC_interaction}

Neutrinos from the \nuc{51}{Cr} source have enough energy to undergo a charged current inverse beta-decay interaction with one of the isotopes of germanium, \nuc{73}{Ge}. This reaction has an energy threshold of $341\,\mathrm{keV}$ and can be expressed as:
\begin{equation}
^{73}\mathrm{Ge} + \nu_e \rightarrow\ ^{73}\mathrm{As} + e^-
\label{eqn:CC_reactionGe}
\end{equation}
The electron will carry the balance of the energy from the neutrino, which is 
$405.7\,\mathrm{keV}$ in the case of the $746.6\,\mathrm{keV}$ neutrino. This provides a unique signature for this reaction, especially given the excellent energy resolution of HPGe detectors. The subsequent \nuc{73}{As} decay has a 80~day half-life and has associated K and L-shell lines that can be used as a consistency check in enriched detectors. There is also a 10\% branching ratio in \nuc{73}{As} decay for a coincident emission of a $53\,\mathrm{keV}$ gamma-ray that can be used as an additional consistency tag.
\\
Cross-section estimates for this reaction do not exist to the authors' knowledge. However, a cross-section of $5.5\times10^{-45}\,\mathrm{cm}^2$ has been computed for a similar process using \nuc{51}{Cr} neutrinos as it relates to SAGE and GALLEX~\cite{bahc95, hata95, haxt98}:
\begin{equation}
^{71}\mathrm{Ga} + \nu_e \rightarrow\ ^{71}\mathrm{Ge} + e^-
\label{eqn:CC_reactionGa}
\end{equation}
This cross-section was also experimentally confirmed by the SAGE experiment~\cite{abdu99}. 
\\
\nuc{73}{Ge} has a natural abundance of 7.73\%. The abundance of \nuc{73}{Ge} in germanium enriched to 85\% in \nuc{76}{Ge} varies significantly, though, from 0.05\% to 1.36\% ~\cite{bara10}. 
Using a cross-section of $10^{-44}\,\mathrm{cm}^2$ and an optimistic abundance of 1\% of \nuc{73}{Ge} in the enriched germanium envisioned for the tonne-scale experiment, one estimates that a tonne-scale Ge detector will detect about one reaction (eqn.~\ref{eqn:CC_reactionGe}) a day from a $10\,\mathrm{MCi}$ \nuc{51}{Cr} source located at a distance of $50\,\mathrm{cm}$. Such a rate at a specific energy should be easily detectable above background. This reaction is sensitive to neutrino flavor and could potentially be used to search for sterile neutrino oscillations, similar to what was proposed in~\cite{grie07}. This will be a topic of future study.  

\section{Conclusions and Outlook}
\label{se:conclusions}
We have estimated the sensitivity of the \MJ\ \DEM\ to measuring the magnetic dipole moment of the neutrino using a $1\,\mathrm{MCi}$ \nuc{51}{Cr} source. The estimates show that the \DEM\ will be competitive, and the results are summarized in table~\ref{tab:money_table}. We discussed what could be done to improve the sensitivity of the experiment beyond the simple analysis methods applied in this paper. We also briefly reviewed the potential physics applications of the charged current reaction, as given in equation~\ref{eqn:CC_reactionGe}. A realistic, intense tritium source was shown to produce a rate below the detectable limit of even a tonne-scale HPGe experiment and is concluded to be not viable for the \MJ\ experiment.

\begin{acknowledgments}
The authors would like to thank Mike Marino for providing the fit parameters to the low-energy part of the \cogent\ spectrum, as well as Juan Collar, Jason Detwiler, Steve Elliott, and John Wilkerson for useful discussions. This work is funded by DOE Office of Nuclear Physics grant number DE-FG02-97ER41041.
\end{acknowledgments}

\bibliography{main}

\end{document}